\documentclass[a4paper,12pt]{article}

\usepackage[a4paper]{geometry}
\usepackage[english]{babel}
\usepackage{amsmath}
\usepackage{amsfonts}
\usepackage{amstext}
\usepackage{amssymb}
\usepackage{amsthm}
\usepackage{amscd}
\usepackage{mathrsfs}
\usepackage[pagebackref,draft=false]{hyperref}
\hypersetup{colorlinks,
linkcolor=myrefcolor,
citecolor=mycitecolor,
urlcolor=myurlcolor}

\usepackage[capitalize]{cleveref}
\usepackage{cite}
\usepackage{caption}
\usepackage{etaremune}

\usepackage{xcolor}
\definecolor{myurlcolor}{rgb}{0,0,0.4}
\definecolor{mycitecolor}{rgb}{0,0.5,0}
\definecolor{myrefcolor}{rgb}{0.5,0,0}
\usepackage{graphicx}
\usepackage{tikz}
\usepackage{tikz-cd}
\usetikzlibrary{cd}
\usetikzlibrary{decorations.pathmorphing,shapes}

\usepackage{etoolbox}
\usepackage{enumitem} 
\usepackage[]{latexsym}
\usepackage{braket}
\usepackage{caption}
\usepackage[utf8]{inputenx}
\usepackage[T1]{fontenc}
\usepackage{lmodern}
\usepackage{textcomp}
\usepackage{microtype}
\usepackage{totcount}
\usepackage{blindtext}
\newtheorem*{proof*}{Proof}

\newcommand{\be}{\begin{equation}}
\newcommand{\ee}{\end{equation}}
\newcommand{\bea}{\begin{eqnarray}}
\newcommand{\eea}{\end{eqnarray}}

\numberwithin{equation}{section}
\numberwithin{theorem}{section}

\title{Feynman's Propagator in Schwinger's picture of Quantum Mechanics}
\date{}

\author{F. M. Ciaglia$^{1,6}$ \href{https://orcid.org/0000-0002-8987-1181}{\includegraphics[scale=0.7]{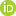}}, F. Di Cosmo$^{1,2,7}$ \href{https://orcid.org/0000-0003-0256-5913}{\includegraphics[scale=0.7]{ORCID.png}}, A. Ibort$^{1,2,8}$ \href{https://orcid.org/0000-0002-0580-5858}{\includegraphics[scale=0.7]{ORCID.png}}, \\ G. Marmo$^{3,4,9}$ \href{https://orcid.org/0000-0003-2662-2193}{\includegraphics[scale=0.7]{ORCID.png}}, L. Schiavone$^{1,3,5,10}$  \href{https://orcid.org/0000-0002-1817-5752}{\includegraphics[scale=0.7]{ORCID.png}}, A. Zampini$^{3,5,11}$ \href{https://orcid.org/0000-0003-0980-6003}{\includegraphics[scale=0.7]{ORCID.png}} \\
\footnotesize{$^{1}$\textit{Depto. de Matem\'aticas, Univ. Carlos III de Madrid, Legan\'es, Madrid, Spain}} \\
\footnotesize{$^{2}$\textit{ ICMAT, Instituto de Ciencias Matem\'{a}ticas (CSIC-UAM-UC3M-UCM)}}\\
\footnotesize{$^{3}$\textit{ INFN-Sezione di Napoli, Naples, Italy}} \\
\footnotesize{$^{4}$\textit{ Dipartimento di Fisica ``E. Pancini'', Universit\`a di Napoli Federico II,  Naples, Italy}} \\
\footnotesize{$^{5}$\textit{ Dipartimento di Matematica e Applicazioni "Renato Caccioppoli", Università di Napoli Federico II, Napoli, Italy}} \\
\footnotesize{$^{6}$\textit{ e-mail: \texttt{fciaglia[at]math.uc3m.es}}} \\
\footnotesize{$^{7}$\textit{ e-mail: \texttt{fcosmo[at]math.uc3m.es}}} \\
\footnotesize{$^{8}$\textit{ e-mail: \texttt{albertoi[at]math.uc3m.es}}} \\
\footnotesize{$^{9}$\textit{ e-mail: \texttt{marmo[at]na.infn.it}}} \\ 
\footnotesize{$^{10}$\textit{ e-mail: \texttt{luca.schiavone[at]unina.it}}}\\
\footnotesize{$^{11}$\textit{ e-mail: \texttt{azampini[at]na.infn.it}}}  
}

\begin{document}

\maketitle

\begin{abstract}
A novel derivation of Feynman's sum-over-histories construction of the quantum propagator using the groupoidal description of Schwinger picture of Quantum Mechanics is presented.  It is shown that such construction corresponds to the GNS representation of a natural family of states called Dirac-Feynman-Schwinger (DFS) states.  Such states are obtained from a q-Lagrangian function $\ell$ on the groupoid of configurations of the system.   The groupoid of histories of the system is constructed and the q-Lagrangian $\ell$ allow to define a DFS state on the algebra of the groupoid.  The particular instance of the groupoid of pairs of a Riemannian manifold serves to illustrate Feynman's original derivation of the propagator for a point particle described by a classical Lagrangian $L$.
\end{abstract}

\section{Introduction:  Feynman versus Schwinger quantum dynamics}	

A new answer to Dirac's query on the role of the Lagrangian in Quantum Mechanics \cite{Di33} that uses in a crucial way the groupoidal formulation of Schwinger's symbolic description of Quantum Mechanics will be discussed.   

Recently it has been shown how the introduction of a q-Lagrangian function\cite{Ci21a} $\ell$ on the groupoid of configurations $K$ of a quantum system allows us to recover a c-Lagragian function $\mathcal{L}$ on the Lie algebroid of the given system that leads to a classical Lagrangian function $L$ providing the standard mechanical description of the system on the tangent bundle of ``velocity'' configurations of the system.     

In this Letter we will elaborate further these ideas by revisiting the answers provided respectively by R. Feynman\cite{Fe05} and J. Schwinger\cite{Sc91} to Dirac's insight.    Both Feynman and Schwinger used their answers to Dirac's question as the fulcrum of their own interpretation of Quantum Mechanics.   

Feynman kept the classical Lagrangian function  $L$  of the theory but proposed a new revolutionary dynamical principle, departing completely from the classical theory, that asserts that the quantum propagator of the theory is given by the well-known Feynman's\cite{Fe48} sum-over-histories formula:
\begin{equation}\label{eq:feynman}
\varphi^F_{(x_1,t_1; x_0,t_0)} = \int_{\Omega (x_0,t_0;x_1,t_1)} \mathcal{D} \gamma \, \,e^{\frac{i}{\hbar} \int_{t_0}^{t_!} L(\dot{\gamma}(s)) \, ds} \, , 
\end{equation}
where $\Omega(x_0,t_0;x_1,t_1)$ is the collection of all paths $\gamma \colon [t_0,t_1] \to \Omega$ joining $x_0$ and $x_1$, and ``$\int \mathcal{D}\gamma$'' represents an \textit{ad hoc} integration technique over histories (the so called `path integral'\footnote{In concrete applications perturbative methods, for instance Feynman diagrammatic techniques, are used to compute the propagator.}).  The previous formula will be called Feynman's dynamical principle and its analysis will constitute the main subject of this work.  

Schwinger, however, took quite a different road.  He postulated a quantum dynamical principle\cite[Chap. 3]{Sc91}, which is inspired on the standard Hamilton's principle of the calculus of variations in classical mechanics, but applied to a quantum Lagrangian $\mathbf{L}$ that turns out to be an operator-valued distribution depending on the quantum fields of the theory.  The amplitudes of the theory, called `transition functions', must satisfy a differential relation that, in the case of quantum mechanical systems, take the simple form\footnote{The general expression for a covariant relativistic theory was given in \cite{Sc51} .}:
\begin{equation}\label{eq:schwinger}
\delta \varphi_{(x_1,t_1; x_0,t_0)} = i \langle x_1,t_1 | \, \delta \int_{t_0}^{t_1} \mathbf{L}(s) ds \, | x_0, t_0 \rangle = i \langle x_1,t_1 | G_1 - G_0  | x_0, t_0 \rangle \, ,
\end{equation}
with $G_1,G_0$ Hermitean operators acting on the Hilbert spaces at $t_1,t_2$ respectively.
The quantum Lagrangian operator $\mathbf{L}$ is guessed by the theoretician and the variations in the previous formula must be computed over all significant variables of the theory.   

A natural question raises immediately, how two such disparate principles could agree in their predictions\footnote{It could be argued that both formulations are equivalent because they lead to the same predictions, however the foundations of both approaches are really different and there is no, as far as we know, a unified conceptual framework for both of them in spite of arguments like in \cite{Sc05} that merely reproduce Schwinger's arguments.}?    In the following paragraphs we will try to argue that the groupoidal description of Quantum Mechanics provides a natural setting for Feynman's principle  by exhibiting a general expression for the quantum propagator of a quantum system in terms of the geometrical properties of the groupoid of configurations of the system and a chosen q-Lagrangian function $\ell$ on it.   It will be shown how this approach applies to the particular simple case of free motion on a Riemann manifold, recovering Feynman's formula (\ref{eq:feynman}).    

The analysis of Schwinger's dynamical principle, Eq. (\ref{eq:schwinger}), from the groupoidal perspective and the relation of Schwinger's quantum Lagrangian $\mathbf{L}$ with the groupoidal q-Lagrangian $\ell$, will be deferred to subsequent work where other applications, like the role of the topology of the underlying manifold in the dynamics will be discussed further.

%%%%%%%%%%%%%%%%
%%%%%%%%%%%%%%%%

\section{Quantum systems and groupoids}

In this section a succinct review of the notations and terminology used in the groupoidal description of Schwinger's picture of Quantum Mechanics will be provided.  As it has been discussed in previous works (see, for instance, Refs.\cite{Ci19,Ci19a,Ci19b,Ci19c,Ci20a,Ci20b}), the departing point for the groupoidal picture of Quantum Mechanics is that the study of a quantum system starts by fixing a certain groupoid $K \rightrightarrows \Omega$, called the groupoid of configurations of the system, that provides the kinematical setting of the system, and that for the purposes of this Letter will be assume to be a Lie groupoid\cite{Ma05}. 

The elements of the space of objects $\Omega$ of the groupoid represent the \textit{outcomes}, or observed quantities, of the theory and are denoted as $x,y,a,b,\ldots$.   The morphisms of the groupoid $\alpha \colon x \to y$, are called \textit{transitions} and provide an abstraction of the notion of ``transition'' or ``relations among observed quantities'' in old quantum mechanics or, quoting Heisenberg\cite{He25} : \textit{``The present paper seeks to establish a basis for theoretical quantum mechanics founded exclusively upon relationships between quantitites which in principle are observable''}.   Transitions, as morphisms of a groupoid, also abstract Schwinger's notion of selective measurements (see for instance \cite{Ci19} and \cite{Ci19a}).  The outcome $x$ of the transition $\alpha \colon x \to y$, will be called the source of $\alpha$ while $y$ will be called its target and denoted respectively by $s(\alpha) = x$ and $t(\alpha ) = y$.

Let $G \rightrightarrows G^0$ be a groupoid with space of objects $\Omega = G^0$, we will denote by $G^x$ the set of transitions whose target is $x\in G^0$ (analogously we denote by $G_x$ the set of transitions whose source is $x$).    Two transitions $\alpha$ and $\beta$ will be said to be composable if $s(\alpha) =t (\beta)$ and their composition will be denoted by $\alpha \circ \beta$.  Units in the groupoid $G$ will be denoted as $1_x \colon x \to x$ and they satisfy $\alpha \circ 1_x = \alpha$ and $1_y \circ \alpha = \alpha$, provided that $\alpha \colon x \to y$. Finally there is an inverse operation $\tau \colon \alpha \mapsto \alpha^{-1}$ such that $\alpha^{-1} \circ \alpha = 1_x$ and $\alpha \circ \alpha^{-1} = 1_y$.

Some basic examples of groupoids are provided by the following ones. Firstly, the groupoid  of pairs $P(\Omega) = \Omega \times \Omega \rightrightarrows \Omega$ of an arbitrary space $\Omega$: it has source and target maps $s(y,x) = x$, $t(y,x) = y$, respectively, composition law $(z,y) \circ (y,x) = (z,x)$, units $1_x = (x,x)$ and inverse $(y,x)^{-1} = (x,y)$. Secondly, standard groups are groupoids, i.e., if $\Gamma$ is a group we consider the groupoid with just one outcome (corresponding to the neutral element) and whose transitions are the elements $g$ of $\Gamma$, the composition law being the composition law in the group $\Gamma$.   

The direct product of the groupoid of pairs $P(\Omega)$ with a group $\Gamma$ provides a new groupoid $G = P(\Omega) \times \Gamma \rightrightarrows \Omega$, whose transitions have the form $\alpha = (y; g; x) \colon x \to y$, $x,y \in \Omega$, $g \in \Gamma$, with composition law $(z; g' ; y) \circ (y; g; x) = (z; g'g; x)$.  

The observables of the theory describing a given system are the real elements of the von Neumann algebra of its groupoid of configurations $K \rightrightarrows \Omega$.    The construction of the von Neumann algebra $\nu(K)$ of the groupoid $K$ proceeds by introducing a theory of integration compatible with the groupoid structure.  This can be accomplished using the general abstract framework provided by A. Connes' non-commutative theory of integration\cite{Co78,Ka82}, or extending Borel's theory of integration on groups to the groupoid situation as done, for instance, by G.W. Mackey, P. Hahn, etc. (see, for instance, Refs. \cite{Ha78,Ha78b}).   Sufficient to say that we will assume that there is a measure class $[\nu]$ defined on a Borel algebra of measurable subsets of the groupoid $K$, and a system of conditional measures $\nu^y$ with support on the fibres $K^y$ of the target map $t \colon K \to \Omega$, such that $\int_K f(\alpha ) \, \mathrm{d}\nu (\alpha) = \int_\Omega \mathrm{d}\nu_\Omega (y) \int_{K^y} \mathrm{d}\nu^y (\alpha) f(\alpha)$ for any integrable function $f$ on $K$. 

The convolution algebra of the measure groupoid $(K, [\nu])$ is the algebra $C_\nu (K)$ of integrable functions $f$ on $K$ such that the convolution product:
$$
(f \star g) (\alpha) = \int_{K^y} \mathrm{d} \nu^y (\gamma) \, f(\gamma) g(\gamma^{-1}\circ \alpha ) \, ,
$$
is well defined. The convolution product $\star$ together with the natural involution $f \mapsto f^*$, 
$$
f^*(\alpha)  = \overline{f(\alpha^{-1})} \Delta(\alpha^{-1}) \, ,
$$ 
with $\Delta \colon K \to \mathbb{R}$ the modular function of the groupoid\footnote{That is, $\Delta^{-1} = \delta \nu^{-1}/\delta \nu$, is the Radon-Nikodym derivative of the measure $\nu^{-1}$ obtained by pushing forward the measure $\nu$ via the inversion $\tau$, with respect to $\nu$.}, makes $C_\nu (K)$ into a $*$-algebra that is represented naturally in the $C^*$-algebra of bounded operators on $L^2(K,\nu)$ by means of its left regular representation $\lambda \colon C_\nu (K) \to \mathcal{B}(L^2(K, \nu))$, that is:
$$
\lambda (f) \Psi = f \star \Psi \, , \qquad \Psi \in L^2(K,\nu) \, .
$$   
The von Neumann algebra $\nu (K)$ of the groupoid is the von Neumann algebra generated by the family of operators $\lambda(f)$, $f\in C_\nu (K)$, i.e., $\nu (K) = \lambda (C_\nu (K))''$, with $\mathcal{S}'$ denoting the commutant of the set $\mathcal{S}$, i.e, the set of all bounded operators $A$ on the given Hilbert space such that  $[A,B] = 0$ for all $B \in \mathcal{S}$, and $(\cdot)''$, the double commutant, that is, the commutant of the commutant.

The states of the theory will be the (mathematical) states on the von Neumann algebra $\nu(K)$, that is, a state is a normalised positive linear functional $\rho \colon \nu (K) \to \mathbb{C}$, $\rho (\mathbf{1}) = 1$, and $\rho (a^*a) \geq 0$ for all $a\in \nu (K)$.   A convenient way to describe a large family of normal states on $\nu (K)$ is by means of functions of positive type on the groupoid $K$.   We will say that a function $\varphi \colon K \to \mathbb{C}$ is of positive type if (see \cite{Ci21} and references therein):
\begin{equation}\label{eq:positive}
\int_K \varphi (\alpha) (f^*\star f)(\alpha) \, \mathrm{d}\nu (\alpha ) \geq 0 \, , \qquad \forall f \in C_\nu(K) \, .
\end{equation}
It will also be assumed that $\varphi$ is normalised, even if this is not strictly necessary for the purposes of this paper, that is, 
$$
\int_\Omega \varphi_{\Omega} (x) \, \mathrm{d}\nu_\Omega (x) = 1\,,
$$
where $\varphi_{\Omega}(x)= \varphi(1_x)$.   Thus, given a normalised function of positive type $\varphi$, we can define a normal state\cite{Bla06} on $\nu(K)$ by the formula:
\begin{equation}\label{eq:state}
\rho_\varphi (f) = \int_K \varphi (\alpha) f(\alpha) \, \mathrm{d}\nu (\alpha) \, .
\end{equation}

%%%%%%%%%%%%%%%%
%%%%%%%%%%%%%%%%

\section{Dirac-Feynman-Schwinger states}\label{sec:Dirac}

For the purpose of obtaining an interpretation of Feynman's principle, Eq. (\ref{eq:feynman}), we will consider a histories-based\cite{So11} approach to our quantum system:  \textit{``...some of us regard histories-based formulations of quantum theories as more basic and more satisfactory than operator formulations, both for the purposes of quantum gravity and for the sake of philosophical understanding''}.  This approach has been considered repeatedly in the standard description of Quantum Mechanics and has led to the development of the notion of decoherence functionals and quantum measures (see for instance Refs.\cite{Ge90,So94}, and references therein).
Thus, it will be assumed that starting from a groupoid of configurations $K \rightrightarrows \Omega$, there is another groupoid, $G \rightrightarrows G^0 = \widetilde{\Omega}$, that will represent a histories-based description of the system (see below, Sect. \ref{sec:histories}, for the construction of the groupoid of histories of a given groupoid).  

The key observation for the remaining discussion is that there is a distinguished family of states, called in what follows Dirac-Feynman-Schwinger (DFS) states, defined by functions of positive type on the groupoid $G$, recall (\ref{eq:positive}), of the form: 
\begin{equation}\label{eq:DF}
\varphi ( w ) = \sqrt{p(s(w )) p(t(w))} \, \, e^{i \mathscr{S}( w )} \, , \qquad w \in G \, ,
\end{equation}
where $\mathscr{S}\colon G \to \mathbb{R}$ is a log-like function, i.e., 
\begin{equation}\label{eq:log}
\mathrm{(a)\,\,} \mathscr{S} (w \circ w') = \mathscr{S}(w) + \mathscr{S} (w') \, ,\qquad \mathrm{(b)\,\,} \mathscr{S} (w^{-1}) = - \mathscr{S} (w) \, , 
\end{equation}
for all composable $w, w'$ in $G$, and the function $p \colon G^0 \to \mathbb{R}$ is a probability density on the space of units of the groupoid $G$.
The functions $\varphi$ defined by (\ref{eq:DF}) satisfies some remarkable properties that can be checked easily:
\begin{enumerate}
\item  Reality: $\varphi^*(w) =  \varphi (w)$, and $|\varphi (w)|^2 = \varphi (w) \varphi (w^{-1}) = p(a)p(b)$, $w \colon a \to b$, i.e.,  $|\varphi (w)|$ depends only on the outcomes $a,b$ of the transition $w$ and the ``quantumness'' of the state is fully encoded on  its phase. 
\item Factorizability:  $\varphi (w) \varphi (w') = p(c) \varphi (w \circ w')$, $c = s(w) = t(w')$.
\item Positivity (Eq. (\ref{eq:positive})):  
\begin{equation}\label{eq:positivity}
 \int_{G} \varphi (w) (f^*\star f)(w) \, \mathrm{d} \nu (w) = || \varphi_f ||_{L^2(G^0)}^2 \geq 0\, ,
 \end{equation} 
 where:
$$
\varphi_f (a) = \int_{G^a} \mathrm{d} \nu^a (w) \,  f(w) \sqrt{p(s(w))} \,  e^{i \mathscr{S}(w)} \, ,
$$
\item Reproducing kernel:  Consider the distribution $\varphi_{ba}$ on $G$ defined as:
\begin{equation}\label{eq:propagator}
\varphi_{ba} = \int_{G_a^b} \mathrm{d} \nu^b (w) \, \sqrt{p(b)p(a)} \, e^{i \mathscr{S}(w)} \, ,
\end{equation}
where $G_a^b$ denotes the collection of transitions $w \colon a \to b$. It is straightforward to show that $\varphi_{ba}$, satisfies the reproducing kernel property:
$$
\varphi_{ba} = \int_{G^c} \mathrm{d}\mu (c) \,  \varphi_{bc} \, \varphi_{ca} \, .
$$
\end{enumerate}

The positivity property (\ref{eq:positivity}) shows that the GNS representation $\pi_\varphi$ provided by the state $\rho_\varphi$ associated to the function $\varphi$ has support on the Hilbert space $\mathcal{H}_\varphi = L^2 (G^0)$, and is given by the formula:
$$
\pi_\varphi (g) \varphi_f = \varphi_{g\star f} \, .
$$
Note that the cyclic vector generating the GNS representation is given by $|0 \rangle = \varphi_{\mathbf{1}}$, where $\mathbf{1}$ is the unit of the von Neumann algebra of the groupoid $G$.  

%%%%%%%%%%%%%%%%
%%%%%%%%%%%%%%%%

\section{The groupoid of histories of a quantum system}\label{sec:histories}

In this section we will discuss how to associate various groupoids of histories to a given quantum system.

The notion of history is the ``continuous'' analogue of the notion of ``chains'' of transitions or  consistent parametrised sequences of groupoid morphisms.   More precisely, given a groupoid $K \rightrightarrows \Omega$, a (future oriented) \textit{chain} on $K$ is a finite sequence of consistent transitions $\alpha_{n_1} , \alpha_{n_1-1}, \cdots, \alpha_{n_0+1}, \alpha_{n_0}$, with $n_0 \leq n_1,$ two integer numbers, and \textit{consistent} means that $s(\alpha_k) = t (\alpha_{k-1})$, $k = n_0 + 1, \ldots, n_1$.  Each one of the transitions $\alpha_k$ in the chain $\{ \alpha_k \}$, will be called a \textit{link}.   

For each chain  $\alpha_{n_1} , \alpha_{n_1-1}, \cdots, \alpha_{n_0+1}, \alpha_{n_0}$, we can define a map $w \colon [n_0,n_1]\subset \mathbb{Z} \to K$,  as:
\begin{equation}\label{eq:wk}
w(k) = \alpha_k \circ \alpha_{k-1} \circ \cdots \circ \alpha_{n_0+1} \circ \alpha_{n_0} \, ,
\end{equation}
where $[n_0,n_1] \subset \mathbb{Z}$ denotes the interval with endpoints $n_0$, $n_1$, that is, $[n_0,n_1] = \{ k \in \mathbb{Z} \mid n_0 \leq k \leq n_1\}$.   In particular $w(n_0) = \alpha_{n_0}$.  We will say that a chain is normalised if the first link is a unit, $w(n_0) = \alpha_{n_0} = 1_{x_0}$. We can always consider the chain to be normalised, by adding an extra link at the beginning if necessary.  We will call the interval $[n_0,n_1]$ the domain of the (parametrised, future oriented) \textit{history} $w$.  Note that given $w$ we recover the sequence of links $\alpha_k$,  by means of the formula:
\begin{equation}\label{eq:alphak}
\alpha_k = w(k) \circ w({k-1})^{-1} \, ,  \qquad n_0 < k \leq n_1 \, .
\end{equation} 
We call the pair $(x_0, n_0) = (s(w(n_0)), n_0)$ the source, denoted by $s(w)$, of the history $w$ and $(x_1,n_1) = (t(w(n_1)), n_1)$, the target, denoted  $t(w)$, of the history $w$.  We will say that the length of the history $w$ is $n_1 - n_0 +1$, i.e., the number of links of its associated chain.

We can define a canonical composition law of histories: $w_2 \circ w_1 \colon [n_0, n_2] \to K$, where $w_1$ and $w_2$ have domains $[n_0,n_1]$ and $[n_1,n_2]$ respectively, and $s(w_2) = t(w_1)$, by means of: 
\begin{equation}\label{eq:comp}
w_2 \circ w_1 (k) = \left\{ \begin{array}{ll} w_1 (k) \, , & \mathrm{if\, } n_0 \leq k \leq n_1 \, ,  \\  w_2(k) \circ w_1(n_1)\, , & \mathrm{ if\, } n_1 \leq k \leq n_2 \, . \end{array} \right.
\end{equation}
   Clearly, the composition $w_2 \circ w_1$ corresponds to the consistent sequence of transitions $\alpha^{(2)}_{n_2}, \ldots, \alpha^{(2)}_{n_1+1} , \alpha^{(2)}_{n_1}\circ \alpha^{(1)}_{n_1} , \ldots, \alpha^{(1)}_{n_0}$, where the links $\alpha^{(a)}_l$, are the links of the history $w_a$, $a = 1,2$, respectively.  
   
The family of (normalised, future oriented) histories satisfies the axioms of a category, i.e., the composition law  is associative: $w_3\circ (w_2 \circ w_1) = (w_3\circ w_2) \circ w_1$ provided that $s(w_3) = t(w_2)$ and $s(w_2) =t(w_1)$, and there are unit elements that correspond to trivial histories which are given by the maps $1_{x,n} \colon \{n\} = [n,n] \to K$, $1_{x,n} (n) = 1_x$.  Then, clearly, $w \circ 1_{x_0,n_0} = w$, provided that $s(w) = (x_0,n_0)$, and $1_{x_1,n_1}\circ w = w$ if $t(w) = (x_1,n_1)$.   

The groupoidification of such category gives us the groupoid of histories $\mathscr{K}$ of the groupoid $K$ we are looking for.   An explicit way of doing it is by considering the free product\cite{Ci20a} of the groupoid $K\times \mathbf{T}_2$ with itself, where $\mathbf{T}_2$ is the groupoid with two elements $\{ P, F \}$ ($P$ stands for ``past'' and $F$ stands for ``future''), and transitions $P \to F$ and $F \to P$, along the family of maps $\sigma^\pm_k \colon \{ P, F\} \to \mathbb{Z}$, given by $\sigma^+_k (P) = k$, and $\sigma^+_k (F) =  k+1$, $k \in \mathbb{Z}$ (future oriented histories), and $\sigma^-_k \colon \{P,F\} \to \mathbb{Z}$, given by $\sigma^-_k (P) = k+1$, and $\sigma^-_k (F) =  k$, $k \in \mathbb{Z}$ (past oriented histories).    With these conventions and notations, the groupoid of histories $\mathscr{K}$ can be written as $\mathscr{K} = \star_{k \in \mathbb{Z}} (K \times \mathbf{T}_2 (\sigma^\pm_k))$, where $\mathbf{T}_2(\sigma^\pm_k)$ denotes the groupoid $\mathbf{T}_2$ whose space of objects is identified with a subset of $\mathbb{Z}$ by means of the maps $\sigma^\pm_k$ (see \cite{Ci20a} for details on the free product of groupoids construction).   Because of the universal definition of the free product each groupoid $K \times \mathbf{T}_2$ is a subgroupoid of $\mathscr{K}$ and
there is a natural filtration of the groupoid of histories $\mathscr{K}$ by the spaces $K^n$ of histories of length $n$, that is $K^0 = \Omega \subset K^1 = K \subset K^2 \subset \cdots \subset K^n \subset \cdots \subset \mathscr{K}$.  

Appealing as it is, we will not pursue this approach to the construction of the groupoid of histories of a groupoid further and we will concentrate on a different structural aspect of the notion of histories that, together with the insight gained from the discrete-time discussion above, will provide the right setting to frame Feynman's sum-over-histories principle in the groupoidal picture.    

Note first that a history $w$ defines a groupoid homomorphism $w \colon P([n_0,n_1]) \to K$, where $P([n_0,n_1])  = [n_0,n_1] \times [n_0,n_1]$ is the groupoid of pairs of integer numbers in the interval determined by $n_0$ and $n_1$, by means of the formula (recall Eqs. (\ref{eq:wk}) and (\ref{eq:alphak})):
\begin{equation}\label{eq:fact}
w(l,k) = w(l) \circ w(k)^{-1} = \alpha_l \circ \alpha_{l-1} \circ \ldots \circ \alpha_{k+1} \, , \qquad n_0 \leq k \leq l \leq n_1\, .
\end{equation}
Note that, clearly, $w(l,k) = w(l,j) \circ w(j,k)$, and $w(k,k) = 1_{x_k}$.    The map $w(\cdot) \colon [n_0,n_1] \to K$ that defines the history $w$, is now recovered as $w(k) = w(k,n_0)$.    Note also that a groupoid homomorphism $w \colon P([n_0,n_1]) \to K$ can always be considered to be the restriction of a groupoid homomorphism $\tilde{w} \colon P(\mathbb{Z}) \to K$ to the subgroupoid $P([n_0,n_1]) \subset P(\mathbb{Z})$, for instance, just taking $\tilde{w}$ to be the constant extension of $w$ from the interval $[n_0,n_1]$ to $\mathbb{Z}$, i.e., $\tilde{w}(k) = w(n_1)$, if $n_1\leq k$, and $\tilde{w}(k) = 1_{x_0}$, if $k \leq n_0$.   

The previous observations give us the hint to introduce the main notion we will use in what follows.  A universal history, or a ``universe'', is a homomorphism of groupoids $w \colon P(\mathbb{R}) \to K$, with $P(\mathbb{R}) = \mathbb{R}\times \mathbb{R}$ the groupoid of pairs $(t,s)$ of real numbers\footnote{In case the groupoid $K$ would carry a measurable, topological or smooth structure, we would like to consider homomorphisms $w$ that will respect such structures.  Thus it would be assumed that $w$ is a measurable map.}, that is, a \textit{universe} $w$ will map every $t \in \mathbb{R}$ into an element $x_t = x(t) \in \Omega$, and any pair $(t,s) \in P(\mathbb{R})$, will be mapped into a transition $w(t,s) \colon x_t \to x_s$, such that: 
\begin{equation}\label{eq:hom}
w(t,s) = w(t,u) \circ w(u,s) \, , \qquad w (t,t) = 1_{x(t)} \, ,\qquad  \forall t,s,u \in \mathbb{R} \, .
\end{equation} 
The label $t$ used to parametrise universal histories is a real number which can be interpreted as the time as measured in the experimental setting provided by the groupoid of configurations $K$, we may call the groupoid of pairs $P(\mathbb{R}) = \mathbb{R}\times \mathbb{R} = \mathbf{T}$, the \textit{Time Groupoid} and be denoted consistently as $\mathbf{T} \rightrightarrows \mathbb{R}$ with target and source maps $t^{(T)}(t',s') = t'$, $s^{(T)}(t',s') = s'$, respectively.   Given a universal history $w \colon \mathbf{T} \to K$, we will call the  associated map, $x \colon [t_0,t_1] \to \Omega$, given by $x(s) = t(w(s,t_0))$, the path (of outcomes on $\Omega$) associated to $w$.  Note that the assignment $w\colon (t,s) \mapsto w(t,s)$, is defined for both $t \leq s$ and $s \leq t$.  In fact, because of the homomorphism property, 
\begin{equation}\label{eq:inverse}
w(t,s)^{-1} = w(s,t) \, , \qquad \forall t,s \in \mathbb{R} \, ,
\end{equation}
 and the notion of the inverse of a history as a history of the system travelling backwards in time is automatically implemented in the formalism.    Finally, similar to formula (\ref{eq:fact}), the homomorphism property (\ref{eq:hom}) allows us to write $w(t,s) = w(t,\tau_0) \circ w (\tau_0, s) = w (t, \tau_0) \circ w(s, \tau_0)^{-1} = w_{\tau_0}(t) \circ w_{\tau_0}(s)^{-1}$, where $w_{\tau_0}(t) =w (t,\tau_0)$, $\tau_0$ being an arbitrary reference time.   Hence, in order to define a universe $w$, it suffices, fixed a reference time $\tau_0$, a map $w_{\tau_0} \colon \mathbb{R} \to K$, and then $w (t,s)$ will be given by the previous formula: 
 \begin{equation}\label{eq:factor}
 w(t,s) = w_{\tau_0}(t) \circ w_{\tau_0}(s)^{-1} \, ,
 \end{equation}
 satisfying, obviously, (\ref{eq:hom}).  Such a map $w_{\tau_0}$ will be called a $K$-path \cite{Cr02} and it satisfies $w_{\tau_0}(\tau_0) = 1_{x_0}$, and $t(w_{\tau_0}(s)) = x_s$, that is, a $K$-path is a lifting to $K$ of a path $s \mapsto x_s$ in $\Omega$ beginning at $x_0 = x_{\tau_0}$.

The groupoid of pairs $P([t_0,t_1]) = [t_0,t_1] \times [t_0,t_1]$ corresponding to any closed interval $[t_0,t_1]\subset \mathbb{R}$ can be considered as a subgroupoid of the Time Groupoid $\mathbf{T}$.  Hence any universal history $w \colon \mathbf{T} \to K$, can be restricted to $P([t_0,t_1])$, and it will determine in this way a groupoid homomorphism $w_{t_1,t_0} \colon P([t_0,t_1]) \to K$, $w_{t_1,t_0} (t,s) = w(t,s)$, for any $t,s$ in the interval determined by $t_0$ and $t_1$.  Note that both situations $t_0 \leq t_1$, and $t_1 \leq t_0$ can happen,  then if $t_0 \leq t_1$ we will say that the time of the concrete history $w_{t_1,t_0}$ flows to the future and, conversely, if $t_1 \leq t_0$, we will say that the time of $w$ flows to the past.  

The restriction of a universe $w$ to a subgroupoid $P([t_0,t_1])$ will be called a concrete history (or just a ``history'' for short) and the interval determined by $t_0$ and $t_1$, its domain.    Moreover if $w_{t_1,t_0}$ is a groupoid homomorphism $w_{t_1,t_0} \colon P([t_0,t_1]) \to K$, there is always an extension of $w_{t_1,t_0} $ to a universal history whose restriction to $P([t_0,t_1]) $ is $w_{t_1,t_0}$ (it suffices to consider the trivial constant extension of $w_{t_1,t_0}$).  Finally, we can associate to any concrete history $w_{t_1,t_0}$ its source $(x_{t_0}, t_0)$ and its target $(x_{t_1}, t_1)$ in $\Omega \times \mathbb{R}$, that is, we define the source and target maps:
 \begin{equation}\label{eq:source}
 s(w_{t_1,t_0}) = (x_0,t_0) \, , \qquad t(w_{t_1,t_0}) = (x_1, t_1) \, . 
 \end{equation}
 As in the case of universal histories, a concrete history is completely determined by fixing a reference time $\tau_0$ in its domain and a map $w_{\tau_0} \colon [t_0,t_1] \to K$, using (\ref{eq:factor}).   Usually we will use the initial time $t_0$ as a reference time, then the concrete history $w_{t_1,t_0}$ will be written as $w_{t_1,t_0}(t,s) = w_{t_1,t_0}(t,t_0) \circ w_{t_1,t_0}(s,t_0)^{-1}$ or,  avoiding unnecessary indices and using the $K$-path notation above, we can write $w_{t_1,t_0}(t,t_0)$ just as $w_{t_0}(t)$, and we obtain again $w_{t_1,t_0} (t,s) = w_{t_0}(t) \circ w_{t_0}(s)^{-1}$.
Note that if we would use another reference time $\tau$, then $w_\tau (t) = w_{t_1,t_0}(t,\tau) = w_{t_1,t_0}(t,t_0) \circ w_{t_1,t_0}(t_0,\tau)$, and we get the following change of reference time formula: 
\begin{equation}\label{eq:change}
w_\tau (t) = w_{t_0}(t) \circ w(t_0,\tau) = w_{t_0}(t) \circ w_{t_0}(\tau)^{-1} \, .
\end{equation}
  
Now, as in the case of discrete-time histories discussed above, concrete histories can be composed in a natural way provided that they are consistent, that is, the target of the first match the source of the second.  Thus if $w^{(1)}_{t_1,t_0}$, and  $w^{(2)}_{t_2,t_1}$ are two concrete histories, they could be composed if $x^{(2)}_1 = w^{(2)}(t_1) = w^{(1)}(t_1) = x^{(1)}_1$. Moreover, mimicking (\ref{eq:comp}), if $t_0 \leq t_1 \leq t_2$, i.e., the time parameter flows to the future, we can define the composition $w^{(2)}_{t_2,t_1}\circ w^{(1)}_{t_1,t_0}$ of the two concrete histories $w^{(2)}_{t_2,t_1}$ and $w^{(1)}_{t_1,t_0}$, as the concrete history with domain $[t_0,t_2]$, reference time $t_0$, determined by the map $(w^{(2)}\circ w^{(1)})_{t_0} (s)$, given by:
\begin{equation}\label{eq:composition}
(w^{(2)} \circ w^{(1)})_{t_0} (s) = \left\{ \begin{array}{ll} w^{(1)}_{t_0} (s) \, , & \mathrm{if\, } t_0 \leq s\leq t_1 \, ,  \\  w_{t_1}^{(2)}(s) \circ w_{t_0}^{(1)}(t_1)\, , & \mathrm{ if\, } t_1 \leq  s \leq t_2 \, ,
\end{array} \right.
\end{equation}
The previous formula can be interpreted naturally in terms of $K$-paths, that is, the composition of two $K$-paths is just the concatenation of the corresponding paths and the composition of their liftings.    Note that the restriction of the composition of the histories $w^{(2)}\circ w^{(1)}$ to the interval $[t_0,t_1]$ is just $w^{(1)}$.  Moreover if we write the $K$-path of $w^{(2)}\circ w^{(1)}$ with respect to the reference time $t_1$, the change of reference time formula (\ref{eq:change}) will give, $t_1 \leq s \leq t_2$:
\begin{equation}\label{eq:restriction}
(w^{(2)}\circ w^{(1)})_{t_1}(s) = (w^{(2)}\circ w^{(1)})_{t_0}(s) \circ (w^{(2)}\circ w^{(1)})_{t_0}(t_1)^{-1} =  w_{t_1}^{(2)}(s) \, ,
\end{equation}
that can be interpreted saying that the restriction of the composition $w^{(2)}\circ w^{(1)}$ to the domain $[t_1,t_2]$ of $w^{(2)}$ is $w^{(2)}$ again.

Clearly the composition of future oriented histories defined by (\ref{eq:composition}) is associative and there are units defined by the trivial histories, i.e, the paths $1_{x_0,t_0}$ that take the value $1_{x_0}$ at $t_0$. Hence the family of future-oriented histories form a category.  

The groupoid of histories $\mathscr{K}$ of a groupoid $K \rightrightarrows \Omega$ can be defined as the groupoid generated by the category of future-oriented histories on $K$.   The inverse of the future-oriented history $w_{t_1,t_0}$ is the past-oriented history $w^{-1}_{t_0,t_1} (s,t) = w_{t_1,t_0}(t,s)$, recall (\ref{eq:inverse}).    The $K$-path associated to the inverse history $w^{-1}$ with respect to the reference time $t_0$ will be given by:
\begin{equation}\label{eq:inversepath}
(w^{-1})_{t_0}(s) = w^{-1}_{t_0,t_1}(s,t_0) = w_{t_1,t_0}(t_0,s) = w_{t_1,t_0}(s,t_0)^{-1} = w_{t_0}(s)^{-1} \, .
\end{equation}
Notice that this is consistent with the composition (\ref{eq:composition}), in fact, we get that $(w^{-1}\circ w)_{t_0}(s) = w^{-1}_{t_0}(s) \circ w_{t_0}(s) = 1_{x_0}$, thus, we get the constant history at $x_0$ (which is the natural extension of $1_{x_0,t_0}$).  

The groupoid of histories $\mathscr{K}$ is a groupoid with space of outcomes $\Omega \times \mathbb{R}$, its source and target maps given by (\ref{eq:source}).    We can denote by $\mathscr{K}_+$ the (sub)category of concrete histories whose time flows to the future and, similarly $\mathscr{K}_-$, the (sub)category of histories whose time flows to the past and $\mathscr{K}_- =  \mathscr{K}_+^{-1}$.  In what follows we will just call an element $w$ in the groupoid $\mathscr{K}$ a \textit{history} (or an \textit{oriented history} if it is the restriction of a universal history)  and it will be denoted consistently as $w \colon (x_0,t_0) \to (x_1,t_1)$.   The space of histories whose target (source) is $(x_1,t_1)$ (resp. $(x_0,t_0)$) will be denoted as $\mathscr{K}^{x_1,t_1}$ (resp. $\mathscr{K}_{x_0,t_0}$), and $\mathscr{K}(x_1,t_1;x_0,t_0)$ the space of histories $w \colon (x_0,t_0) \to (x_1,t_1)$, with given source $(x_0,t_0)$ and target $(x_1,t_1)$.

A compact way of expressing the defining property of the groupoid of histories is by writting $\mathscr{K} =  \mathscr{K}_+ \star\mathscr{K}_-$, where $\star$ denotes the free product of categories, that is, the category consisting on finite reduced words with alphabet $\mathscr{K}_\pm$, and characterised by the following universal property: Let $C_l$ be a family of categories and $\iota_l \colon C_l \to \star_l C_l$ be the natural identification $\iota_l (w_l) = w_l$, $w_l \in C_l$, and let $\phi_l \colon C_l \to C$ be injective functors from the categories $C_l$ to the category $C$, then there exists a functor $\Phi \colon \star_l C_l \to C$ such that $\Phi \circ \iota_l = \phi_l$. 

Note that the composition given by formula  (\ref{eq:composition}) serves as well if the time of both histories flows to the past, however two concrete histories (corresponding to two different universes $w^{(1)}$, $w^{(2)}$) with different time flows should be considered as independent regarding the free product, that is, its composition is not going to be the concrete history of an universe.   However the composition of two histories with different time flows can be interpreted nicely as the concatenation of the corresponding $K$-paths, that is, if $w^{(1)}$ flows forward in time with domain $[t_0,t_1]$ ($t_0 \leq t_1$) and $w^{(2)}$ flows backwards in time with domain $[t_2,t_1]$ (i.e, $t_2 \leq t_1$), then the composition $w^{(2)}\circ w^{(1)}$ is given by the $K$-path $(w^{(2)} \circ w^{(1)})_{t_0} \colon [t_0,t_1] \sqcup [t_2,t_1] \to K$, with $[t_0,t_1] \sqcup [t_2,t_1]$ denoting the disjoint union of the intervals $[t_0,t_1]$ and $[t_2,t_1]$, and:
$$
(w^{(2)} \circ w^{(1)})_{t_0} (s) = \left\{ \begin{array}{ll} w^{(1)}_{t_0} (s) \, , & \mathrm{if\, } s \in [t_0,t_1] \, ,  \\  w_{t_1}^{(2)}(s) \circ w_{t_0}^{(1)}(t_1)\, , & \mathrm{ if\, } s \in[t_2, t_1] \, .
\end{array} \right.
$$

If the groupoid of configurations is a Lie groupoid, the notion of histories so far formulated allows us to recover the notion of links that served so well in the discrete-time case.   If $K \rightrightarrows \Omega$ is a Lie groupoid its infinitesimal description is provided by its associated Lie algebroid\cite{Cr03} $\pi\colon A(K) \to \Omega$.   Let us recall that the Lie algebroid $A(K)$ of the Lie groupoid $K$ is the family $\xi_x$ of values at units $x\in \Omega$ of right-invariant vector fields $X^\xi$ on $K$.  The anchor map $\mu \colon A(K) \to T\Omega$ of the Lie algebroid $A(K)$ is defined as $t_*(\xi_x)$, with $t\colon K \to \Omega$ the target map of $K$.  In the particular instance of the groupoid of pairs $P(\Omega)$ of a manifold, its Lie algebroid $A(P(\Omega))$ is just the tangent bundle $T\Omega \to \Omega$ with anchor map the identity map $\mu = \mathrm{id} \colon T\Omega \to T\Omega$.   

Given a Lie algebroid $\pi \colon A \to \Omega$ with anchor map $\mu \colon A \to T\Omega$, an $A$-path\cite{Cr02,Cr03} is a smooth map $\xi \colon [t_0,t_1] \to A$, such that $\mu (\xi (s)) = \frac{d}{ds} \pi (\xi(s))$.   An $A$-path $\xi (s)$ can also be thought as a Lie algebroid homomorphism $\boldsymbol{\xi} \colon T\mathbb{R} \to A$,  where $T\mathbb{R}$ is the Lie algebroid of the groupoid of pairs $\mathbb{R}\times\mathbb{R}$, the relation between both notions given by: $\boldsymbol{\xi} (s,\frac{d}{dt}) = \xi (s)$.   

Because a history $w$ is as a groupoid homomorphism $w \colon \mathbf{T} \to K$, it induces, provided that it is differentiable, a homomorphism of Lie algebroids $w_* \colon T\mathbb{R} \to A(K)$, by means of $w_*(s,d/dt) = TR_{w(s)^{-1}} \dot{w}(s)$, where $R_w$ denotes the right translation by $w$ on the groupoid of histories and $\dot{w}$ is a tangent vector to $\mathscr{K}_w$ (see, for instance, Ref. \cite{Ma05} for details), or in a more familiar notation reminiscent of the corresponding formula for Lie groups: 
\begin{equation}\label{eq:paths}
\xi (s) = \dot{w}(s) \circ w(s)^{-1} \, .
\end{equation}
Obviously, the correspondence $w\mapsto \xi$, between universal histories $w$ and $A(K)$-paths $\xi$ can be restricted to any subgroupoid $P([t_0,t_1])$ of $\mathbf{T}$, that is, there is a natural correspondence between concrete histories and $A(K)$-paths, with domains arbitrary intervals $[t_0,t_1]\subset \mathbb{R}$, given by formula (\ref{eq:paths}).  If $w$ is not an oriented history, then the corresponding $A(K)$-path should be understood as a juxtaposition of standard $A(K)$-paths with domain a disjoint union of intervals.

The correspondence between histories and $A(K)$-paths in the previous extended sense, $w \in \mathscr{K} \mapsto \xi = \dot{w} \circ w^{-1} \in \mathscr{A}_K$, with $\mathscr{A}_K$ denoting the space of $A(K)$-paths, has a right inverse provided by the exponential map, that is: $\mathrm{Exp}(\xi (s)) = w(s)$.   Note that for a given groupoid the exponential map is defined on sections of the Lie algebroid\cite{Ma05} $A(K)$, that is $\mathrm{Exp}$ is defined on a map $\tilde{\xi} \colon \Omega \to A(K)$ such that $\pi (\tilde{\xi}(x)) = x$; and we would like to define it just for sections $\xi (s)$, along a map $x \colon s \mapsto x(s)$ in $\Omega$.   This difficulty is easily solved by considering an arbitrary extension $\tilde{\xi}$ of the section $\xi(s)$ to $\Omega$ and restricting the values of the exponential $\mathrm{Exp}(x, \tilde{\xi}(x))$, to $x(s)$. Such definition will not depend on the chosen extension (note that the exponential is defined as the flow of the right-invariant vector field defined by $\xi$, thus it depends only on the initial conditions).

Now we are ready to answer the question raised at the end of Section \ref{sec:Dirac}:  Given a groupoid of configurations $K$ describing a quantum system:  How do we construct a Dirac-Feynman-Schwinger state?

%%%%%%%%%%%%%%%%
%%%%%%%%%%%%%%%%

\section{Feynman meets Schwinger}\label{sec:Feyman_groupoid}

A real Lagrangian function $\ell$ on the groupoid of configurations $K$ provides the key ingredient to construct a DFS state on the groupoid of histories $\mathscr{K}$ of a quantum system.    We will call such function the quantum Lagrangian, or the q-Lagrangian, of the theory (see \cite{Ci21a} for additional context).  Given a  q-Lagrangian $\ell \colon K \to \mathbb{R}$, there is a natural Dirac-Feynman-Schwinger state $\varphi_\ell$ defined on the von Neumann algebra of the groupoid of histories $\mathscr{K}$, given by the action functional: 
$$
\mathscr{S}(w) = \int_{t_0}^{t_1} \ell (w_{t_0}(s)) \, \mathrm{d} s = \int_{t_0}^{t_1} \ell (w(s,t_0))\, \mathrm{d} s \, ,
$$ 
for $w_{t_0} (s)$ the $K$-path of a future oriented history.   The corresponding definition of the action functional for a past oriented history $\tilde{w}$ will be:
$$
\mathscr{S}(\tilde{w}) =  -\int_{t_0}^{t_1} \ell (\tilde{w}_{t_0}(s))\, \mathrm{d}s \, ,
$$
hence, clearly, because (\ref{eq:inversepath})
$$
\mathscr{S}(w^{-1}) =  -\int_{t_0}^{t_1} \ell (w^{-1}_{t_0}(s)) \, \mathrm{d} s = -\int_{t_0}^{t_1} \ell (w_{t_0}(s)^{-1}) \, \mathrm{d} s  =  -\int_{t_0}^{t_1} \ell (w(t_0,s)) \, \mathrm{d} s   \, .
$$
Then, if $ \ell (\alpha ) = \ell (\alpha^{-1})$, condition (\ref{eq:log}b) is satisfied:
$$
\mathscr{S}(w^{-1})  = - \mathscr{S}(w) \, .
$$
Moreover, note that for $w_1, w_2$, two composable future-oriented histories with domains $[t_0,t_1]$ and $[t_1,t_2]$ resp., using (\ref{eq:restriction}), we get:
\begin{eqnarray*}
\mathscr{S}(w_2 \circ w_1) &=&  \int_{t_0}^{t_2} \ell ((w_2 \circ w_1)(s,t_0)) \, \mathrm{d} s  \\ &=&  \int_{t_0}^{t_1} \ell ( (w_2 \circ w_1)(s,t_0)) \, \mathrm{d} s +  \int_{t_1}^{t_2} \ell ( (w_2 \circ w_1)(s,t_1)) \, \mathrm{d} s \\ &=&   
  \int_{t_0}^{t_1} \ell ( w_1 (s,t_0)) \, \mathrm{d} s +  \int_{t_1}^{t_2} \ell ( w_2 (s,t_1)) \, \mathrm{d} s = 
\mathscr{S}(w_2) + \mathscr{S}(w_1) \, ,
\end{eqnarray*}
which gives condition (\ref{eq:log}a).     We conclude, because of the positivity property (\ref{eq:positivity}), that the functional $\varphi^\ell \colon \mathscr{K} \to \mathbb{C}$, defined by means of:
$$
\varphi^\ell (w) = \sqrt{p(s(w)) p(t(w))} e^{i/\hbar \,  \mathscr{S}(w)} \, ,
$$
with $p$ a probability density\footnote{See \cite{Ci20c} for a discussion of the role of classical systems in the groupoidal picture.} on $\Omega \times \mathbb{R}$ and $\hbar$ a constant introduced because of dimensional reasons,
defines a state on the von Neumann algebra of the groupoid of histories of $K$. 

It is noticeable that such DFS state will depend solely on the q-Lagrangian function $\ell$ and the background probability density $p$ (apart from the measure theory used to construct the von Neumann algebra of the groupoid of histories).  We may consider that such density $p$ is absorbed in the construction of the measure $\nu$ on the groupoid $\mathscr{K}$, hence, we can conclude, because of the discussion in Sect. \ref{sec:Dirac}, Eq. (\ref{eq:propagator}), that the propagator of the theory determined by $\ell$ (and $\nu$) will be given by:
\begin{equation}\label{eq:schwinger_propagator}
\varphi^\ell_{(x_1,t_1;x_0,t_0)} = \int_{\mathscr{K}(x_1,t_1;x_0,t_0)}  \, \, e^{i/\hbar \int_{t_0}^{t_1} \ell (w(s)) ds} \, \, \mathrm{d} \nu ( w ) \, , 
\end{equation}
where $(x_1,t_1)$ and $(x_0,t_0)$ are respectively the target and the source of the histories $w \colon (x_0,t_0) \to (x_1,t_1)$ in $\mathscr{K}(x_1,t_1; x_0,t_0)$.    No attempt will be made in this paper to provide an explicit construction of the measure $\nu$ that appears in (\ref{eq:schwinger_propagator}), so that the previous expression should be considered somehow formal, very much as in any treatment of path integral formulas.   Nevertheless, it must be pointed out that the flexible approach to the theory of integration on groupoids provided by Connes's theory of Non-commutative integration \cite{Co78} offers a natural interpretation of the previous formula, aspect that will be discussed elsewhere.

As it was discussed in Ref.\cite{Ci20c} , there is a natural classical system associated to the quantum system with groupoid of configurations $K \rightrightarrows \Omega$.  Such system is determined by the Abelian algebra $L^\infty (\Omega )$, which is just the von Neumann algebra of the totally disconnected subgroupoid defined by the units $1_x$ of the groupoid $K$.   Then, the restriction of the DFS state defined by $\varphi^\ell$ to $L^\infty (\Omega\times \mathbb{R})$, is the measure defined by the function $p$, that is, the restriction of the positive-type function $\varphi^\ell$, Eq. (\ref{eq:DF}), to the subgroupoid of units gives (note that $\mathscr{S}(1_{x_0,t_0}) = 0$) the ``classical'' positive function:
$$
\varphi_{\mathrm{class}}(x,t) = p(x,t) \, .
$$ 

We may use the infinitesimal description of the groupoid $K$, i.e., its Lie algebroid $A(K)$, to construct a different expression for the propagator of the theory Eq. (\ref{eq:schwinger_propagator}).     The exponential map is a right-inverse to the tangent map that maps histories $w$ to $A(K)$-paths, that is, if $\xi_w = \dot{w} = Tw (d/dt)$, then  $w(s) = \mathrm{Exp\,} (x(s), \xi (s))$.  Then we can push-forward the measure $\nu$ to the space $\mathscr{A}(K)$ of $A(K)$-paths.   Then, we can rewrite the propagator (\ref{eq:schwinger_propagator}) as a Feynman-like propagator:
\begin{eqnarray*}
\varphi_{(x_1,t_1;x_0,t_0)} &=& \int_{\mathscr{K}(x_1,t_1;x_0,t_0)}  \, \, e^{i/\hbar \int_{t_0}^{t_1} \ell (\mathrm{Exp\,} (x,\xi/c)) ds} \, \mathrm{d} \nu (  \mathrm{Exp\,} (x, \xi ) )  \\ &=&  \int_{\mathscr{A}_K(x_1,t_1;x_0,t_0)} \,  e^{i/\hbar \int_{t_0}^{t_1} \mathcal{L}(x(s), \xi (s)) ds} \mathcal{J}(\xi) \,  \mathcal{D} \xi  \, ,
\end{eqnarray*}
where $\mathcal{L}(x,\xi ) = \ell (\mathrm{Exp\,} (x,\xi/c))$, is the c-Lagrangian $\mathcal{L} \colon A(K) \to \mathbb{R}$, associated to the q-Lagrangian $\ell$, $\mathcal{J}(\xi)$ is the Jacobian of the transformation $w \mapsto \xi$,
$\mathscr{A}_K(x_1,t_1;x_0,t_0)$ denotes the space of $A(K)$-paths with endpoints $(x_1,t_1)$ and $(x_0,t_0)$, and $\mathcal{D}\xi$ a measure on the space of $A(K)$-paths induced from a metric on the Lie algebroid $A(K)$.   

The exponential map is differentiable at the space of units with differential the identity map.  Hence, formally, the Jacobian $\mathcal{J}$ can be taken to be trivial.   Moreover assuming that the exponential map defines a Borel isomorphism between the space of $K$-histories and the space of $A(K)$-paths, then we can define a Feynman-like propagator by means of the formula:
$$
\varphi^F_{(x_1,t_1;x_0,t_0)} =  \int_{\mathscr{A}_K(x_1,t_1;x_0,t_0)}  e^{i/\hbar \int_{t_0}^{t_1} \mathcal{L}(x(s), \xi (s)) ds} \,  \mathcal{D} \xi \, ,
$$   
that gives as a particular instance Feynman's celebrated formula (\ref{eq:feynman}).

Thus, a particular application of the previous analysis is Feynman's original sum-over-histories computation of the quantum propagator.  If we consider the groupoid of pairs $P(\Omega)$ of a Riemannian manifold, the space of histories on $[t_0,t_1]$ can be identified with the space of usual paths on $\Omega$. Actually given a history $w \colon [t_0,t_1] \to \Omega \times \Omega$ with source $(x_0,t_0)$, its $K$-path has the form $w_{t_0}(s) = (x(s), x_0)$, with $w_{t_0}(t_0) = 1_{x_0} = (x_0,x_0)$.  The natural q-Lagrangian $\ell$ on the groupoid $P(\Omega)$ is given by the energy function:
$$
\ell (x_1,x_0) =\inf_{\gamma \colon x_0 \to x_1} \frac{1}{2}\int_{t_0}^{t_1} || \dot{\gamma}(s) ||^2 ds \, ,
$$
 Then the formula for the Feynman propagator becomes:
\begin{equation}\label{eq:feynman_propagator}
\varphi^F_{(x_1,t_1;x_0,t_0)}  = \int_{\Omega (x_1,t_1; x_0,t_0)}  \, e^{\frac{i}{\hbar} \int_{t_0}^{t_1} \frac{1}{2} m || \dot {\gamma}(s) ||^2 ds} \mathcal{D} \gamma \, ,
\end{equation}
where we have used the fourth order approximation to the c-Lagrangian $\mathcal{L}(x, v ) = mc^2 \ell ( \exp (v(c))$ computed in Ref.\cite{Ci21a}.

A few comments are in order here.   The ``measure'' $\mathcal{D} \gamma$ can be considered as a cylindrical measure determined by the family of projections $\mathrm{ev}_P \colon \mathcal{K} \to \Omega^{|P|}$, where $P = \{ t_0 = s_0 < s_1 < \cdots < s_r =  t_1 \}$, is an arbitrary finite partition of the interval $[t_0,t_1]$, $|P| = r$, and we consider the natural product measure on $\Omega^{|P|}$ induced by the volume of the Riemannian structure on $\Omega$.     Such choice, as it was commented before, will provide a kernel on the groupoid of histories rather than an actual measure, but this is sufficient to develop the theory of integration required to make sense of the previous formulas.

Another interesting situation that allows to test our conclusions consists of the quantum dynamical description of a system defined by a group $\Gamma$.  There are two ways to address such situation, one in which the elements of the group represent the transitions of the system, that is, the group is the isotropy group of a groupoid with just one object, or alternatively, we may consider that the group itself is the space of outcomes of the system.   That is, consider for instance a particle moving on a circle $S$ which is identified with the group $U(1)$.   This situation becomes another instance of the discussion above where the Riemannian manifold $Q$ is replaced by the group $\Gamma$ equipped with the Killing-Cartan form of the group.  Such situation has been widely discussed in the literature and explicit expressions for the Feynman propagator (\ref{eq:feynman_propagator}) can be found (see, for instance \cite{Ma79}).    

%%%%%%%%%%%%%%%%
%%%%%%%%%%%%%%%%

\section{Conclusions and discussion}

A direct route from Schwinger's conceptual foundation of Quantum Mechanics to Feynman's sum-over-histories principle is provided whose key ingredients are the groupoidal abstract description of Schwinger's algebra of selective measurements, its groupoid of histories and the construction of a natural class of states, called DFS states, determined by a q-Lagrangian function $\ell$ satisfying a ``time reversal'' invariance condition, on the groupoid of configurations of the system.  

From these premises a natural representation of the propagator of the theory  is obtained.   Such q-Lagrangian function can be considered as a real element of the von Neumann algebra of observables of the system, thus it provides a natural meaning to Schwinger's quantum Lagrangian operator $\mathbf{L}$.  The choice of a q-Lagrangian function $\ell$ (together with a classical probability distribution $p$) determines a DFS state on the groupoid of histories of the system, whose GNS representation provides a representation of the propagator of the theory as a sum-over-histories formula, deriving in this way a general form for Feynman's dynamical principle.
Then, we conclude that the groupoidal description of quantum systems provides a common ground to analyse both Feynman's principle and Schwinger's quantum dynamical principle, thus allowing to compare them, task that will be done elsewhere.

A simple familiy of examples have been succinctly discussed, that of free motion of point particles with no internal structure, that amounts to the discussion of the fundamental q-Lagrangian studied in Ref.\cite{Ci21a}.   The standard path integral formulas describing the dynamics are easily derived.   In forthcoming papers systems with inner degrees of freedom will be studied and the role played by the topology of the underlying space of outcomes will be investigated.

%%%%%%%%%%%%%%%
%%%%%%%%%%%%%%%

\section*{Acknowledgments}

The authors acknowledge financial support from the Spanish Ministry of Economy and Competitiveness, through the Severo Ochoa Programme for Centres of Excellence in RD (SEV-2015/0554).
AI would like to thank partial support provided by the MINECO research project  MTM2017-84098-P  and QUITEMAD++, S2018/TCS-A4342. GM would like to thank partial financial support provided by the Santander/UC3M Excellence  Chair Program 2019/2020, and he is also a member of the Gruppo Nazionale di Fisica Matematica (INDAM), Italy. 
F.D.C. thanks the UC3M, the European Commission through the Marie Sklodowska-Curie COFUND Action (H2020-MSCA-COFUND-2017- GA 801538) and Banco Santander for their financial support through the CONEX-Plus Programme.
L.S. would like to thank the support provided by Italian MIUR through the Ph.D. Fellowship at Dipartimento di Matematica R.Caccioppoli.
%%%%%%%%%%%%%%%
%%%%%%%%%%%%%%%

\end{document}